# Agglomeration and filtration of colloidal suspensions with DVLO interactions in simulation and experiment

Bastian Schäfer[a,d], bastian.schaefer@mvm.uka.de, ++49/1637968110

Martin Hecht[b,e], martin.hecht@icp.uni-stuttgart.de, ++49/71168563607

Jens Harting[c,b], j.harting@tue.nl, ++31/402473766

Hermann Nirschl[a], hermann.nirschl@kit.edu, ++49/7216082400

[a] Institut für Mechanische Verfahrenstechnik, Strasse am Forum 8, D-76128 Karlsruhe, Germany

[b] Institut für Computerphysik, Pfaffenwaldring 27, D-70569 Stuttgart, Germany

[c] Technische Natuurkunde, TU Eindhoven, Den Dolech 2, NL 5600 MB Eindhoven, The Netherlands

[d] Bayer Technology Services GmbH, Chempark Geb. E41, 51368 Leverkusen, Germany

[e] High Performance Computing Center (HLRS), University of Stuttgart, Nobelstr. 19, 70569 Stuttgart, Germany

**Graphical abstract**

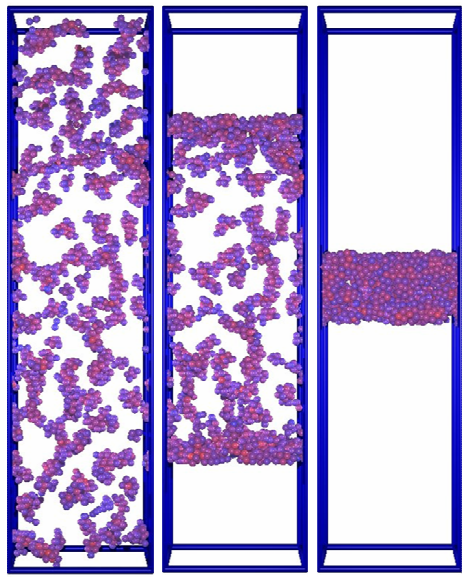

The combination of molecular dynamics (MD), stochastic rotation dynamics (SRD) and lattice Boltzmann (LB) simulations permits to study the agglomeration of colloidal particles (left), the filtration (right) and the permeation of the compressed filter cakes (right).

# 1. Abstract

Cake filtration is a widely used solid-liquid separation process. However, the high flow resistance of the nanoporous filter cake lowers the efficiency of the process significantly. The structure and thus the permeability of the filter cakes depend on the compressive load acting on the particles, the particles size, and the agglomeration of the particles. The latter is determined by the particle charge and the ionic strength of the suspension, as described by the Derjaguin-Landau-Verwey-Overbeek (DLVO) theory. In this paper, we propose a combined stochastic rotation dynamics (SRD) and molecular dynamics (MD) methodology to simulate the cake formation. The simulations give further insight into the dependency of the filter cakes' structure on the agglomeration of the particles, which cannot be accessed experimentally. The permeability, as investigated with lattice Boltzmann (LB) simulations of flow through the discretized cake, depends on the particle size and porosity, and thus on the agglomeration of the particles. Our results agree qualitatively with experimental data obtained from colloidal boehmite suspensions.

# 2. Introduction

Cake filtration is an energy efficient and widely used solid-liquid separation process, where solid particles are retained at a filter medium or membrane and build up a filter cake with increasing thickness. The high flow resistance of nanoporous filter cakes and the growing market for fine particles demand for methods to enhance the filtration, for example by flocculating the suspensions prior to filtration. For improving the filtration processes and apparatuses, a simulation tool is required that comprehends the agglomeration of the suspension, the filtration process, and the pressure-driven permeation of the filter cakes.

Various simulation tools have been applied to different aspects of the filtration process. Some simulations are based on Darcy's law and phenomenological equations for the local porosity and local permeability of the filter cake [1-4]. Kim et al. consider the aggregates as solid cores with porous shells and determine the filter cake's permeability with Stokes' equation and Brinkman's extension of Darcy's law [1,2]. Lao et al. replace the pore system with a network of pipes and junctions and then calculate the flow in the tubes with the Poiseuille equation [3]. Eisfeld et al. propose a pseudo-continuous model for statistically described domain geometries, where the solid/liquid interaction is represented by a coupling term in the Navier-Stokes equation [4]. Solving the Navier-Stokes equation for the complicated geometries of porous filter cakes is very time-consuming for large numbers of particles [5]. The complexity increases further if the particles are mobile [6].

Molecular dynamics (MD) is the standard method for simulating the motion of discrete particles, but the simulation of suspensions is computationally far too demanding if the water molecules and ions are resolved. Even if the water is replaced by a background friction and stochastic fluctuations and only the ions are simulated explicitly, the power of today's computers would limit the simulation to only a few colloidal particles. Omitting the water molecules and ions and assuming that every particle collision leads to agglomeration reduces the computational effort significantly, but this approach does not include electrostatic interactions [7]. Their influence is considered in Monte Carlo studies on the porosity and cake structure of filter cakes [8]. Barcenas et al. present a simple way to control the particle agglomeration in Monte Carlo simulations: the suspension is represented by a two-component mixture of colloidal particles and inhibitor particles [9]. Yu et al. perform Monte Carlo simulations based on the fractal nature of the pore size distribution in porous media [10].

Alternative methods to simulate mobile particles in suspensions became popular in recent years: Brownian dynamics simulations include Brownian motion, but the hydrodynamics is reduced to a simple Stokes force [11-14]. Stokesian dynamics include multiparticle hydrodynamic interactions, but the numerical effort increases with the third power of the particle number [15]. This problem can be reduced by using accelerated Stokesian dynamics [16].

Another popular approach is to combine MD simulations of the solid particles with a simulation of the liquid, for example based on the Navier-Stokes equations, dissipative particle dynamics, the lattice Boltzmann (LB) method or stochastic rotation dynamics (SRD). In these methods, the numerical effort for calculating the hydrodynamic interactions grows linear with the particle number [17]. Dissipative particle dynamics comprise the hydrodynamics and Brownian motion [18]. The lattice Boltzmann method is a powerful tool for modelling single and multiphase flow, which was extended to simulate particle agglomeration including Brownian motion [19, 20]. SRD is an efficient simulation method based on a simple algorithm that includes thermal noise and hydrodynamic interactions of a real fluid [21-24]. The method is also known as multi particle collision dynamics. It has been successfully applied to colloidal suspensions [25-27] and flow in confined geometries [26, 28, 29].

We approach the agglomeration and filtration of colloidal suspensions with a combination of SRD for the electrolyte solution and MD for the colloidal particles. The MD simulation includes the particle-particle interactions as described by the DLVO theory [30, 31]. The permeability of the fixed structure of the filter cake is investigated with the LB method. The methods are explained in detail in chapter 4, including the boundary conditions which represent the filter cell. The article also presents the theoretical background of filter cake formation and permeation in chapter 3. The numerical data are compared to experimental results in chapter 5.

## 3. Theoretical background

Filtration processes are commonly described by Darcy's equation for the flow rate of the filtrate [32-34]

$$\dot{V}_L = \frac{A_C}{\eta} \frac{\Delta p}{R_F}, \qquad (1)$$

with the cross-section area of the filter cake $A_C$, the dynamic viscosity of the permeate $\eta$, and the driving pressure difference $\Delta p$. Neglecting the membrane resistance, the flow resistance

$$R_F = \frac{h_C}{K} \qquad (2)$$

is the ratio of the thickness of the filter cake $h_C$ to its permeability $K$. The permeability mainly depends on the particle size and the porosity of the filter cake $\Phi$, that is the ratio of the liquid volume $V_L$ to the total volume of the filter cake $A_C h_C$. The porosity depends on the particle size, the filtration pressure, and the agglomeration of the particles [35, 36]. According to the well-established DLVO theory, colloidal particles agglomerate due to the van-der-Waals attraction if the Coulomb repulsion is not sufficiently strong to keep them apart. The Coulomb repulsion depends on the ionic strength of the suspension and on the surface charge, as quantified by the zeta potential $\zeta$, which depends on the pH value of the suspension. The material specific pH value where the surface charge equals zero is called isoelectric point (IEP) [37, 38]. There are three regimes of a suspension's stability against agglomeration:

- For a low zeta potential or a high ionic strength, the van-der-Waals attraction is stronger than the Coulomb repulsion at every inter-particle distance. The agglomerated particles are kept apart at very short distances by the Born repulsion.
- If the ionic strength is low and the zeta potential is high or intermediate, the Coulomb repulsion forms an energy barrier against a further approach of the particles. Since the particles' kinetic energy is Boltzmann distributed with an average energy of 1.5 $k_B T$ [39], a range from rapid to slow agglomeration is found for increasing energy barriers. The small secondary energy minimum outside of the energy maximum is too shallow to hold back particles against mechanical load [38, 40].

- For an intermediate ionic strength and a high or intermediate zeta potential, the secondary minimum is deeper than $2\,k_BT$. The particles are captured in the minimum and thus form secondary agglomerates. These are less stable than primary agglomerates because of the significantly smaller energy minimum.

The agglomerate structure can be described with the pair correlation function

$$G(d) = \frac{V}{N^2} \left\langle \sum_n \sum_{n \neq m} \delta\left(d - |x_m - x_n|\right) \right\rangle, \qquad (3)$$

with the examination volume $V$ and the number of particles $N$ in that volume. It gives the probability for the particles to find another particle at a certain center-to-center distance $d$. The peaks of $G$ indicate regular structures, for example a peak at two particle radii originates from particles in direct contact and peaks at larger distances show more complex structures [41]. The filtration behavior, pore structure, and permeability of a filter cake depend significantly on the presence and the size of agglomerates in the suspension. Agglomerated suspensions lead to loosely textured packed beds with large pores between the agglomerates. The agglomerates can be considered as large particles with an internal porosity and so-called macro-pores between the agglomerates which are accountable for the high permeability [34]. In contrast, packed beds formed from stable suspensions with high surface charge have a dense structure with a homogeneous pore size distribution, a low porosity, and a low permeability [42]. The filterability can thus be increased by changing the pH value in the direction towards the IEP or by increasing the ionic strength of the suspension.

Porosity and permeability of nanoporous filter cakes further depend on the filtration pressure [43]. Filter cakes from agglomerated suspensions are highly compressible because the agglomerates can be easily deformed. The rearranging of the particles mainly reduces the size of the large pores between the agglomerates [42, 44, 45], which are accountable for the major part of the fluid transport. Consequently, the compression of the macropores between the agglomerates has a large effect on the permeability, even if a large porosity remains inside the agglomerates [42]. Filter cakes formed from stable suspensions are less compressible because of the lack of interagglomerate pores. As shown by Singh et al., the compressibility of filter cakes from colloidal silica spheres slightly decreases with increasing ionic strength as long as it stays below the critical coagulation concentration [46].

## 4. Simulation methods

For simulating the agglomeration and filtration, the motion of the spherical particles is calculated with a molecular dynamics (MD) simulation. The fluid is simulated with stochastic rotation dynamics (SRD), which is described below. The thermal contribution to the kinetic energy of the MD and SRD particles is controlled with a Monte Carlo thermostat described in [23] to ensure a simulation at constant temperature.

The simulation starts with 2000 MD-particles that are stochastically distributed in the simulation space. To obtain a volume concentration of 3.7%, which is consistent with the experiments, the system dimensions are set to 76.8 particle diameters in the vertical direction of compression and 19.2 particle diameters in the horizontal directions. The boundaries in the horizontal directions are periodic, while the closed boundaries in the vertical direction exert a Hooke force on overlapping particles. Simulations with 4000 particles yield comparable porosities of the packed beds, so that finite-size effects are excluded.

The particles agglomerate until an equilibrium structure is reached (see figure 1, left), as controlled by analyzing the temporal development of the pair correlation function $G$. Subsequently, the filtration takes place by incrementally approaching the z-boundaries of the MD-space towards each other. Consequently, the MD particles build up filter cakes on the top and bottom boundary conditions, like on the membranes in a two-sided filtration apparatus (see figure 1, center). After both filter cakes merge, they are compressed between the membranes (see figure 1, right) until the Hooke force between the z-boundaries and the MD particles reaches the desired compressive load. Like in reality, the compression is carried by the particle network. The fixed positions of the particles define the geometry of the porous structures, which are further examined using the lattice Boltzmann (LB) method.

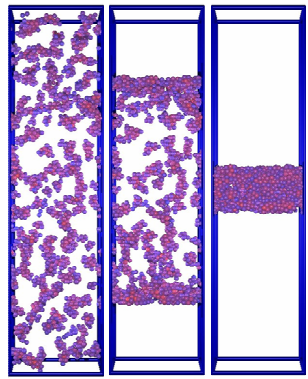

**Figure 1: MD particles in the simulation space at different stages of the simulation (left: agglomeration, center: filtration, right: compression). The SRD particles and the MD boundaries are not visible.**

### *4.1. MD simulation of the particles*

The motion of the particles is calculated with a velocity Verlet algorithm, using Newton's equation of motion. To reduce the computational effort, the MD simulation space is mapped with a cubic grid that has a grid constant of 4 particle radii and the particle interactions are restricted to MD particles in the same and adjacent cells. The interaction force is the derivative of the attractive and repulsive interaction potentials between the particles. For a pair of round particles, the van-der-Waals potential can be written as [40, 47]

$$\Psi_{vdW}(d) = -\frac{A_H}{12}\left[\frac{d_P^2}{d^2 - d_P^2} + \frac{d_P^2}{d^2} + 2\ln\left(\frac{d^2 - d_P^2}{d^2}\right)\right], \quad (4)$$

with the Hamaker constant $A_H$, the particle diameter $d_P$ and the center-to-center distance $d = x_m - x_n$ between particles with positions $x_m$ and $x_n$. The Coulomb potential $\Psi_{Coul}$ for a pair of identically charged spheres is [38, 40]

$$\Psi_{Coul} = \pi\varepsilon_{rel}\varepsilon_0 \left[\frac{2 + d_P}{1 + d_P\kappa} \cdot \frac{4k_BT}{ze}\tanh\left(\frac{ze\zeta}{4k_BT}\right)\right]^2 \frac{(d_P)^2}{d}\exp(-\kappa[d - d_P]), \quad (5)$$

with the vacuum permittivity $\varepsilon_0$, the relative permittivity $\varepsilon_{rel}$, the Boltzmann constant $k_B$, the temperature $T$, the elementary charge $e$ and the valency of the ions $z$. The reciprocal Debye length [38, 40]

$$\kappa = \left(\frac{2F^2 I}{\varepsilon_{rel}\varepsilon_0 N_A k_B T}\right)^{\frac{1}{2}} \quad (6)$$

is a function of the ionic strength $I$, with the Faraday constant $F$ and the Avogadro constant $N_A$. In order to reduce the potential gradients and thus expand the simulation time step, the Born potential, which keeps agglomerated particles from overlapping, is replaced by the less steep Hertz potential

$$\Psi_{Hertz} = K_{Hertz} \cdot (d_P - d)^{2.5}, \quad (7)$$

for $d < d_P$ with the Hertz constant $K_{Hertz} = 0.1$. The particles can consequently overlap to a small extend, but since the overlap is relatively small, its influence on the structure of the filter cake can be neglected. The Hertz force is only relevant for agglomerated particles, since the particles are otherwise prevented from overlapping by the DLVO interactions.

The van-der-Waals potential is cut-off for inter-particle distances below 0.005 radii to circumvent its singularity at direct contact. The gap between the cut-off radius and the particle surface is modeled by a Hooke law with a coefficient $D_H$ of $7.7 \cdot 10^6$ N/m. The coefficient is chosen such that the potential is steadily differentiable at the transition point between the Hooke law and the DLVO potential. The resulting potential in the simulations agrees very well with the real potential in the range of inter-particle distance that determines the agglomeration.

The kinetic energy of approaching particles is partly dissipated due to the viscous behavior of the fluid, which is pressed out of the closing space between the particles. Although SRD is a hydrodynamic simulation method, it does not fully reproduce this behavior for small inter-particle distances because the fluid in the SRD simulation is coarse grained and the algorithm for coupling the MD simulation to the SRD simulation, which is described in chapter 4.4, does not resolve the fluid's viscosity at distances below one particle diameter. However, the algorithm was chosen because of its high numerical efficiency. For particles being closer than one particle diameter, this shortcoming is corrected by the dissipative lubrication force [25]

$$F_{Lub} = -(\dot{x}_m - \dot{x}_n) \frac{6\pi c_{Lub} \eta}{d + r_{CO,i} - d_P} \left(\frac{d_P}{4}\right)^2, \tag{8}$$

with the particle velocities $\dot{x}_m$ and $\dot{x}_n$, the lubrication constant $c_{Lub}$ and inner cut-off radius $r_{CO,i}$ to exclude the singularity for touching particles. The lubrication constant $c_{Lub}$ takes into account that the hydrodynamic interaction is only partly reproduced by the SRD algorithm if the colloidal particles come too close. With a value of $c_{Lub} = 0.2$, the simulation reproduces the expected particle interactions and the stability diagram presented in figure 5.

## *4.2. SRD simulation of the liquid*

The SRD method introduced by Malevanets and Kapral [21] is used for simulating the fluid because it intrinsically contains fluctuations, has low demands for computational time and is applicable to colloidal suspensions [22, 28, 29, 31, 51]. SRD is based on calculating the continuous positions of virtual fluid particles at the time $t + \Delta t_{SRD}$ from the previous positions and velocities. Since the particles are pointlike, they cannot collide. Instead, momentum is exchanged between the SRD particles in a collective interaction step: the SRD particles are sorted into cubic cells with the length $L_{Cell}$. Within each cell $k$, the relative velocities of the particles $\dot{x}_m(t) - \dot{\bar{x}}_k(t)$ with respect to their average velocity $\dot{\bar{x}}_k$ are rotated [48]. The rotation matrix $\Omega_{SRD,k}$ is stochastically chosen for each cell and time step from a set of six possible rotations by +90° or -90° around the three coordinate axes. This is a mathematically simple means to exchange momentum between the fluid particles while conserving the total mass, energy, and momentum within each cell. If the mean velocity is interpreted as the streaming velocity of the fluid in each cell, the relative velocities represent the thermal fluctuations. Since the rotation does not affect the mean velocity of the fluid particles, the total momentum and the kinetic energy of the streaming velocity do not change. Also the kinetic energy associated to thermal fluctuations remains constant, because the rotation of the thermal fluctuations does not influence their magnitude.

If the particle velocities are small, so that most particles remain in their cell, neighboring particles are correlated over several time steps. This correlation is broken if the particles are observed from a moving inertial system with a moving grid. The simulation is thus not Galilean-invariant, which would require that an experiment is independent of the inertial frame from which it is observed. However, in SRD simulations these correlations are minimized if the mean free path of the SRD particles

$$\lambda = \Delta t_{SRD}\sqrt{k_B T / m_{SRD}} \tag{9}$$

is chosen to be at least half the length of the SRD cells, where the mass of one SRD particle is given by

$$m_{SRD} = \frac{L_{Cell}^3 \rho_L}{\langle N_k \rangle} \tag{10}$$

and $\rho_L$ denotes the liquid density [48]. Therefore, under this condition, Galilean invariance is restored. In this study, the average number of SRD particles per cell is chosen as $\langle N_k \rangle = 60$, the length of the SRD cells is given in table 1.

### *4.3. Scaling of the SRD simulation parameters*

The simulation of colloidal suspensions is numerically demanding since the simulation time steps required to resolve the motion of the particles decreases with decreasing particle diameter $d_P$ and due to the increasing complexity of the particle interactions. Therefore, a careful scaling of the physical parameters is applied in order to increase the time steps, while preserving the physical behavior of the system, as explained in what follows. The scaling is an established means to reduce the numerical effort of coarse-graining simulations.

Each force or process in the system takes a characteristic time for translating a particle by the distance of one diameter (see figure 2). For example, the sedimentation time $\tau_{Sed}$ is reciprocal to the Stokes velocity $v_{Sed}$ and the particle diameter $d_P$:

$$\tau_{Sed} = \frac{d_P}{v_{Sed}} = \frac{18\eta}{d_P g (\rho_S - \rho_L)}, \tag{11}$$

with the gravitational constant $g$, the liquid density $\rho_L$, and the particle density $\rho_S$. The diffusion time $\tau_{Diff}$, as calculated from the particles' mean square displacement $\overline{(\Delta x)^2} = 2D\Delta t$, with the diffusion constant $D$ resulting from the Stokes-Einstein relation, is thus proportional to the third power of the diameter [49]:

$$\tau_{Diff} = \frac{d_P^2}{2D} = \frac{3\pi\eta d_P^3}{2k_B T}. \qquad (12)$$

The average velocity of the particles' thermal fluctuations is calculated from their average kinetic energy of $E_{kin} = 0.5 m_m v^2 = 1.5 k_B T$ [50], leading to a characteristic time of

$$\tau_{TF} = \sqrt{\frac{d_P^5 \pi \rho_S}{18 k_B T}}. \qquad (13)$$

The particles' exponential relaxation time

$$\tau_P = \frac{1}{18} \frac{d_P^2 \rho_S}{\eta} \qquad (14)$$

for a particle to adapt to the flow field of the surrounding fluid follows from expanding Newton's equation of motion around a stationary state:

$$\frac{\pi}{6} d_P^3 \rho_S \ddot{x}_m = 3\pi d_P \eta \dot{x}_m. \qquad (15)$$

The diffusive momentum transport in the fluid is described by the fluid molecules' mean square displacement $\overline{(\Delta x)^2} = 2\eta/\rho_L \Delta t$, thus leading to the time scale for the fluid relaxation

$$\tau_L = \frac{d_P^2 \rho_L}{2\eta}. \qquad (16)$$

Since the smallest characteristic time and thus the simulation time steps decrease with decreasing particle diameter, the simulation of colloidal suspensions is limited by the power of today's computers. However, the characteristic times can be expanded while maintaining the physical behavior of the suspension by carefully scaling the physical properties of the system.

The dotted lines in figure 2 refer to the modified time scales after applying the scaling factors for a particle diameter of 30 nm, as explained below. The physical behavior of the system is reproduced if the scaling is done carefully, meaning without changing the sequence of the characteristic times or bringing them close together: Each time scale ratio represents a characteristic number of the system, for example the Reynolds number $Re = \tau_L/\tau_{Sed}$. Increasing the Reynolds number from $2.1\cdot 10^{-11}$ in the original system to $3.8\cdot 10^{-8}$ for the scaled system does not change the physical behavior. The same applies to the Archimedes number $Ar = (9\tau_L)/(2\tau_{Sed})$, the Schmidt number $Sc = \tau_{Diff}/\tau_L$, and the Péclet number $Pe = \tau_{Sed}/\tau_{Diff}$.

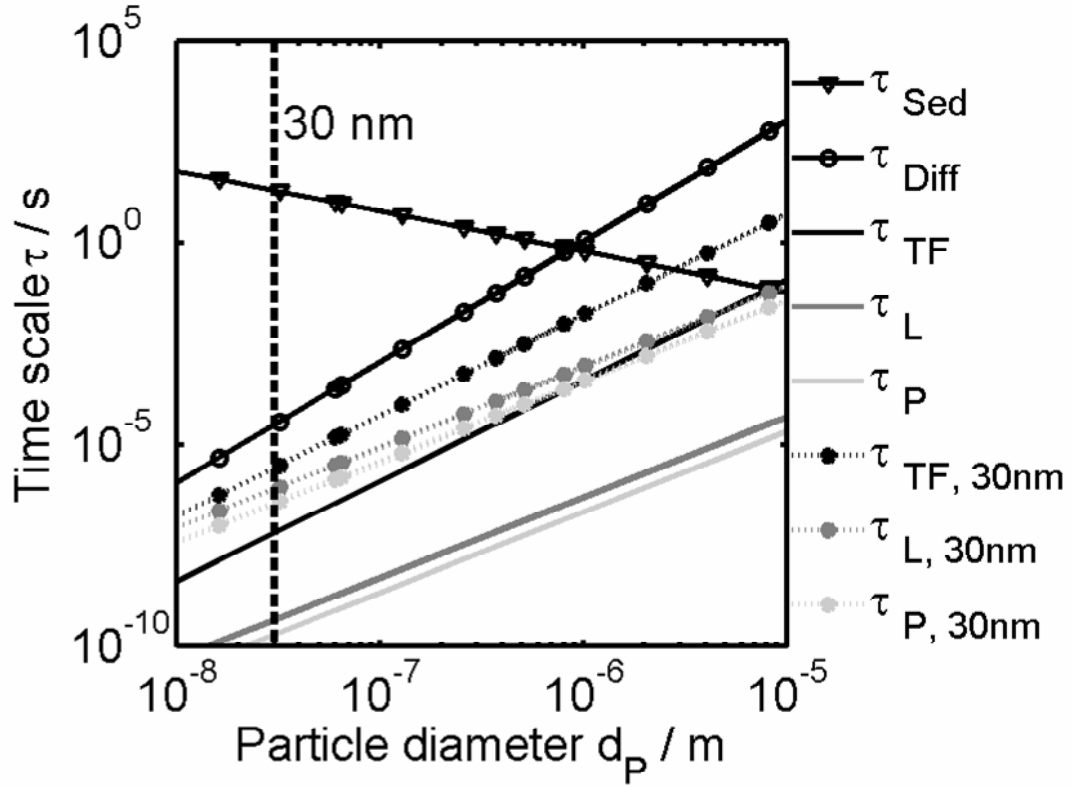

**Figure 2: Characteristic times for different forces in the simulation before scaling (solid lines) and after scaling for a particle diameter of 30 nm (dotted lines). The symbols are explained in the text.**

The SRD simulation used in this study involves a coarse-graining which is accompanied by a scaling of the fluid viscosity. For rotation angles of ±90°, the viscosity of the SRD fluid is

$$\eta_{SRD} = \rho_L \left( \frac{L_{Cell}^2}{18\Delta t_{SRD}} \left(1 - \frac{1-e^{-\langle N_k \rangle}}{\langle N_k \rangle}\right) + \frac{k_B T_{SRD} \Delta t_{SRD}}{4 m_{SRD}} \frac{\langle N_k \rangle + 2}{\langle N_k \rangle - 1} \right), \quad (17)$$

with the temperature of the SRD fluid $T_{SRD}$

$$T_{SRD} = \frac{3\pi \eta_{SRD} d_P D}{k_B} \quad (18)$$

to reproduce the realistic diffusion constant $D$ of the solid particles, as given in table 1. With $\lambda = L_{Cell}/2$ to guarantee Galilean invariance [48], equations (9), (10), (17), and (18) lead to the time steps and SRD viscosity given in table 1. Also the gravitational constant and the inter-particle forces are scaled to reproduce the sedimentation velocity and the ratio of the inter-particle forces to the energy associated to the thermal fluctuations of particles.

**Table 1: Simulation parameters for different particle diameters**

| Particle diameter / nm | 30 | 70 | 120 |
|---|---|---|---|
| Length of the SRD cells / nm | 25 | 50 | 100 |
| Diffusion constant / m²/s | $1.44 \cdot 10^{-11}$ | $6.17 \cdot 10^{-12}$ | $3.60 \cdot 10^{-12}$ |
| SRD viscosity / Pa s | $5.66 \cdot 10^{-7}$ | $2.828 \cdot 10^{-7}$ | $1.414 \cdot 10^{-7}$ |
| SRD time step / s | $1.25 \cdot 10^{-7}$ | $1.10 \cdot 10^{-6}$ | $8.50 \cdot 10^{-6}$ |
| Scaling factor | $5.66 \cdot 10^{-4}$ | $2.828 \cdot 10^{-4}$ | $1.414 \cdot 10^{-4}$ |
| Scaled temperature / K | $1.67 \cdot 10^{-1}$ | $8.34 \cdot 10^{-2}$ | $4.17 \cdot 10^{-2}$ |
| Scaled gravitational constant / m/s² | $5.55 \cdot 10^{-3}$ | $2.77 \cdot 10^{-3}$ | $1.39 \cdot 10^{-3}$ |

This scaling permits to significantly increase the simulation time steps and thus reduce the computational effort for the simulation of colloidal particles, while reproducing the physical behavior of the system.

### *4.4. Coupling of MD and SRD simulations*

The solid particles of the MD simulation are coupled to the SRD simulation by including them in the rotation step. The different masses $m_m$ of the MD and SRD particles are considered as weighting factors when calculating the mean velocities in the cells

$$\bar{\dot{x}}_k(t) = \sum_{m=1}^{N_k(t)} \dot{x}_m(t) m_m \bigg/ \sum_{m=1}^{N_k(t)} m_m, \quad (19)$$

with the number of particles $N_k(t)$ in the cell $k$ [23]. While the SRD particles cover an average distance of $0.5\, L_{Cell}$ during each SRD time step, the thermal fluctuations of the MD particles are much faster. Consequently, smaller time steps are chosen for the MD simulations, namely $1 \cdot 10^{-10}$ for $d_P = 30$ nm, $1 \cdot 10^{-9}$ for $d_P = 70$ nm and $1 \cdot 10^{-9}$ for $d_P = 120$ nm. The SRD-calculation is applied less often than the MD calculation, which reduces the computational effort substantially.

## 4.5. Determination of the permeability by lattice Boltzmann simulations

The lattice Boltzmann (LB) method is a mesoscopic approach to simulate the motion of viscous fluids. The fluid is represented by particle populations located on the nodes of a lattice and the velocity space is discretized to only a few basic lattice vectors $e_j$. Fluid motion is described by the single-particle distribution functions $f_j(x,t)$, which give the expected number densities of particles moving along the lattice vectors $e_j$ for each lattice site $x$ and time $t$. During each time step $\Delta t_{LB}$, the particle distributions are propagated to the next lattice site $x + e_i$, as described by the discretized Boltzmann equation [51, 52]

$$f_j(x + e_j, t + \Delta t_{LB}) = f_j(x,t) + \Delta t_{LB} \Omega_{LB,j}(f(x,t)). \qquad (20)$$

The collision operator $\Omega_{LB,j}$ mimics the viscous behavior of the fluid by relaxing it towards equilibrium.

In the context of this study, the LB simulation can be seen as a Navier-Stokes solver based on a very simple algorithm that is capable to cope with complex flow domains [53-55]. LB simulations are applicable to liquid flow in nanoporous filter cakes since the relevant characteristic numbers, i.e. the Reynolds number $Re$, the Mach number $Ma$ and the Knudsen number $Kn$, are well below one for both the experiment and the simulation. The LB method is ideally suited for simulating fluid flow in porous media since it provides a simple method to represent the complicated pore geometry: the so-called mid-grid bounce back condition is a common way to implement no-slip boundary conditions on the surface of the porous sample [56-58]. Since the boundary conditions are imposed locally, even complex boundaries do not significantly increase the computational time [59]. Furthermore, LB simulations are easy to parallelize because locally only information of the nearest neighbor nodes is required [57].

The fixed geometries of the filter cakes, as obtained from the MD- and SRD-simulation, are mapped on a cubic lattice with 128 nodes in each direction, with the compressed filter cake being centered in the computational domain. The fluid reservoirs above and below the filter cake ensure that the flow field adapts to the complex pore network [51]. The hydraulic force that drives the permeation is applied to a layer of nodes in the inlet reservoir by including an additional force term in the collision operator $\Omega_{LB,j}$ [60]. The permeability of the filter cake is calculated from the difference of the pressure in the inlet and outlet reservoir, calculated as

$$p(x,t) = c_S^2 \rho_{LB}(x,t), \qquad (21)$$

and the fluid's velocity in the direction of flow, averaged over the whole sample.

## 5. Experiments

The experiments are performed in an Electro-Compression-Permeability-Cell, which is described in detail in [61, 62]. The Electro-Compression-Permeability-Cell is designed for the filtration of colloidal suspensions between two membranes with a cross section area of 0.005 m² in a cylindrical tube. The particles in the suspensions agglomerate according to their surface charge and the ionic strength of the suspensions. In analogy to the simulation set-up, the particles are retained by the membranes and build up a filter cake, while the liquid can drain through the filter cake and the membrane. To obtain a homogenous structure, the filter cake is compressed between the membranes by the force of a plunger until the equilibrium thickness between 3 mm and 8 mm is reached. Applying a pressure of 20 kPa to a liquid reservoir on the upstream side of the filter cake drives the liquid through the filter cake onto a scale on the downstream side [61].

For each pH value and ionic strength a new filter cake is formed. The suspensions are prepared by dispersing 25 g of boehmite particles (Disperal®, Disperal 20® or Disperal 40® from Sasol, Germany) in 225g potassium nitrate solution with the desired ionic strength and a pH value of 2.7. The pH value is subsequently adjusted by adding caustic potash. Disperal®, Disperal 20® and Disperal 40® were chosen for the experiments because they have the same chemical composition, but different particle sizes. The number-weighted mean diameter $d_{50,0}$ of the dispersed primary particles is 24 nm for Disperal®, 73 nm for Disperal 20®, and 130 nm for Disperal 40®., as measured with a Nanotrac from Mircotrac Inc., USA. Boehmite has an isoelectric point (IEP) at a pH value of 9.5.

## 6. Results

The simulation methods introduced above are used to investigate the influence of the particle size, the particle charge, and the ionic strength of the suspension on the filtration of suspensions and the permeability of the resulting filter cakes. The colloidal particles in the suspensions agglomerate according to the DLVO theory. Subsequently, the filtration of the suspension is started by reducing the simulation space of the particles. After the filter cakes are compressed to their equilibrium thickness, a lattice Boltzmann simulation yields their permeabilities $K$.

## 6.1. Agglomeration of the particles

The formation of agglomerates, which dominates the filtration behavior of the suspension and the permeability of the filter cakes, depends on whether the kinetic energy of the particles is sufficient to overcome the energy maximum in the DLVO potential. The averaged energy of the thermal degrees of freedom of the simulated particles equals $1.5\, k_B T$ for the three-dimensional simulations, which is consistent with the theory. Furthermore, the viscosity of the liquid, as calculated from the Stokes-Einstein relation

$$\eta = \frac{k_B T}{3\pi d_P D}, \qquad (22)$$

agrees with the viscosity of water. The diffusion constant of the molecular dynamics (MD) particles is calculated from the particle positions.

The agglomeration of the particles is analyzed by means of the pair correlation function $G(d)$, which is shown in figure 3 for the different particle diameters. The peaks refer to the center-to-center distance of the dark spheres in figure 3 . Similar pair correlation functions are observed in Brownian Dynamics simulations [12]. The gap of $G$ between 2 and 2.2 radii indicates the low probability of finding a particle in this steep region of the DLVO potential of another particle. The extension of the first peaks to less than two radii indicates an overlapping of the particles. This results from substituting the Born repulsion by the Hertz force in the DLVO interactions in order to reduce the potential gradient and the computational time. However, the penetration depth due to the overlapping is very small. For large particle distances, the pair correlation function relaxes to one. The large height of the peaks at two particle radii indicates the very narrow distribution of the particle distances for touching particles, which results from the steepness of the particle interaction potentials in the vicinity of the primary minimum.

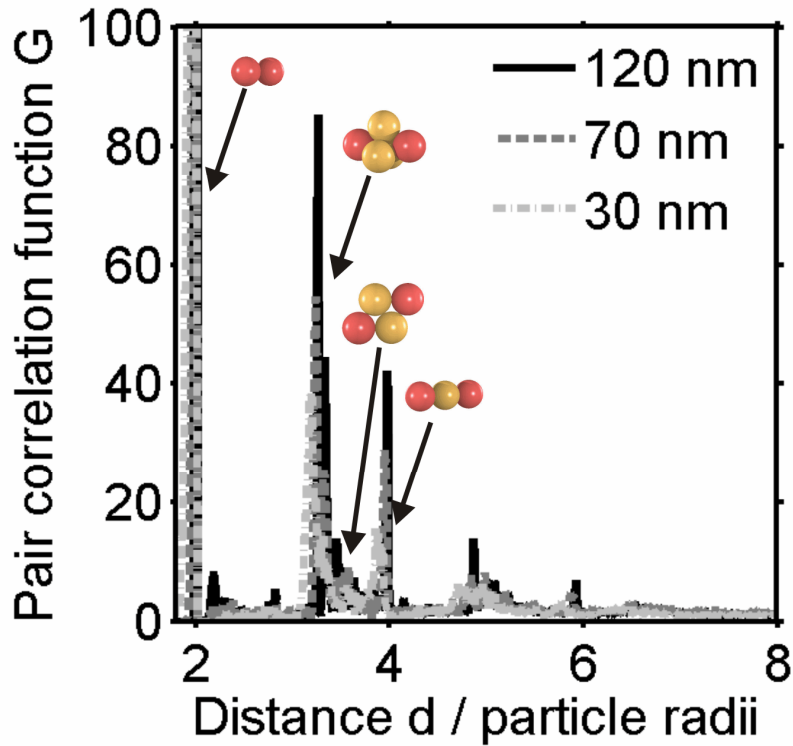

**Figure 3: Pair correlation function $G(d)$ for agglomerated particles of different diameters in the suspension. The sketches show the structures that are represented by the peaks. The peak positions refer to the distance between the dark spheres.**

The positions of the peaks are used to detect primary and secondary agglomeration and to distinguish between them. For secondary agglomerates, the peaks of $G$ are at larger distances than for primary agglomerates (see figure 4) because the secondary minimum is at a larger surface distance. Furthermore, the peaks for secondary agglomerates are significantly broader due to the larger width of the secondary minimum. This indicates a higher mobility of the particles in the agglomerates. Secondary agglomerates are thus less stable against deformation. For secondary agglomerates, the higher order peaks of $G$ are less pronounced.

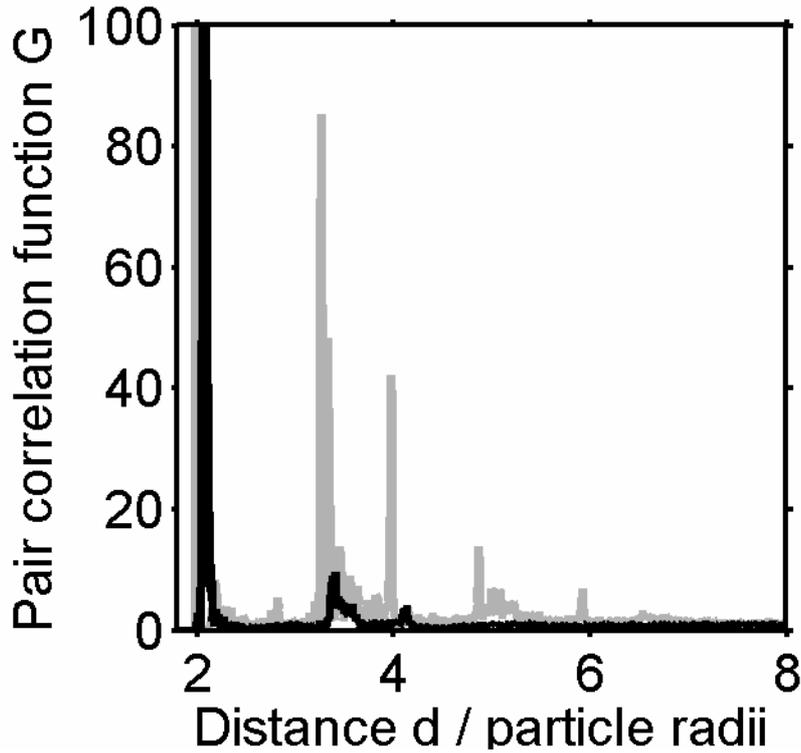

**Figure 4:** Pair correlation functions $G(d)$ for a particle diameter of 120 nm in the presence of primary agglomerates (grey) and secondary agglomerates (black).

The stability diagram (see figure 5) shows the agglomeration behavior of suspensions with different particle diameters depending on the zeta potential $\zeta$ and the ionic strength $I$. For combinations above and between the solid lines, the difference between the maximum and the secondary minimum of the DLVO potential is smaller than 10 $k_BT$. This energy barrier is overcome by at least some of the particles due to the Boltzmann distribution of the kinetic energy [39]. Between the solid lines and the dashed lines, the energy barrier is higher than 10 $k_BT$ and the secondary minimum is deeper than 2 $k_BT$, so that some particles get caught in the secondary minimum. The suspensions are stable below the dashed lines, where the secondary minimum is too shallow to retain the particles. The particle size has a stronger influence on secondary agglomeration than on primary agglomeration.

The symbols show the state of agglomeration observed in the simulations for particle diameters of 30 nm (black symbols), 70 nm (dark grey symbols), and 120 nm (light grey symbols). The data points for a diameter of 70 nm are shifted vertically for the sake of optical clarity, since they would otherwise have the same positions like the data points for 120 nm. Primary agglomerates are indicated by triangles, secondary agglomerates by circles and unagglomerated suspensions by crosses. The agglomeration behavior observed in the simulations agrees with the stability diagram.

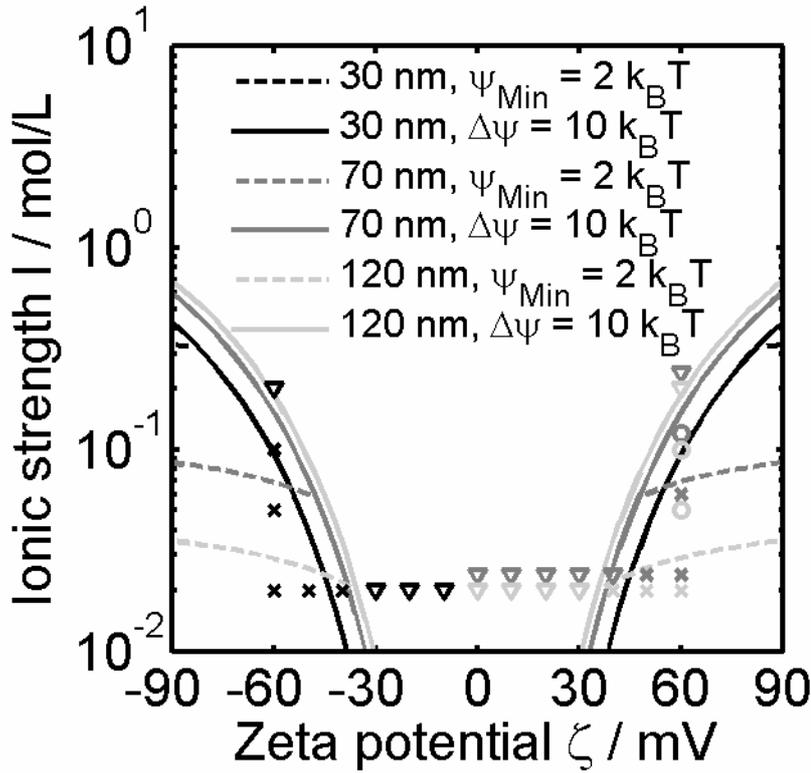

Figure 5: Stability diagram for different particle sizes showing the regions of primary agglomeration (above the solid lines), secondary agglomeration (between the solid and the dashed lines) and stable suspensions (below the dashed lines). The agglomeration behavior observed in the simulations is indicated by triangles (primary agglomerates), circles (secondary agglomerates), and crosses (stable suspensions) for particle diameters of 30 nm (black symbols), 70 nm (dark grey symbols), and 120 nm (light grey symbols). The data points for a particle diameter of 70 nm are shifted vertically for the sake of clarity.

The simulations start with statistically distributed particles which move and agglomerate due to Brownian motion. Before the filtration starts, the agglomeration must come to an equilibrium state, which is controlled via the evolution of the pair correlation function. Beginning with a completely irregular structure, the nearest-neighbor peaks evolve rapidly. The peaks for higher orders follow and reach their equilibrium values after about 0.7 ms for a particle diameter of 30 nm, 4.4 ms for a particle diameter of 70 nm, and 30 ms for a particle diameter of 120 nm. The agglomeration time is longer for larger particles because of the slower Brownian motion, but this also permits larger time steps. The filtration starts directly after the specified agglomeration times since a longer agglomeration would primarily increase the computational time.

### 6.2. Porosity

After the agglomeration reaches its equilibrium, the filtration starts by incrementally reducing the MD-simulation space. This procedure is controlled by the integral force exerted on the boundary planes. Filter cakes build up at the upper and lower boundaries of the MD simulation until they eventually merge and get compressed. The resulting structure is evaluated in terms of its porosity $\Phi$, which is a function of the total volume of the particles and the filter cake

$$\Phi = 1 - \frac{\frac{4}{3}\pi R^3 N_{MD}}{A_C h_C}. \qquad (23)$$

The thickness of the filter cake $h_C$ should not be determined from the coordinates of the highest and lowest particles because the filter cakes are rough and irregular if agglomerates are present. The resulting statistical uncertainty could be reduced by increasing the number of particles, which would also increase the computational effort. Instead, the thickness of the filter cake is calculated from the vertical coordinates $x_{z,m}$ of the particles as

$$h_C = 4 \frac{\sum_{i=1}^{N} \left| x_{z,m} - \frac{\sum_{m=1}^{N} x_{z,m}}{N_{MD}} \right|}{N_{MD}} + d_P. \qquad (24)$$

The porosity of the filter cakes strongly depends on the ionic strength of the suspension (see figure 6), because the ions reduce the Debye length and thus shield the Coulomb repulsion. Accordingly, the porosity is relatively high at an ionic strength of 0.2 mol/L, where the particles are agglomerated. Below 0.1 mol/L, only few agglomerates are formed because of the high energy barrier. The increase of porosity upon a decrease of the ionic strength from 0.1 mol/L to 0,02 mol/L is probably caused by the increasing Debye length: for unagglomerated structures the inter-particle distance depends on the equilibrium of the electrostatic repulsion between the particles and the compressive force on the filter cake. Analogously, the compressibility of filter cakes from colloidal silica spheres decreases with increasing ionic strength as long as it is kept below the critical coagulation concentration [46]. This effect is more pronounced for smaller particles.

A higher pressure during compression results in a lower porosity of the filter cake (see figure 6). The compressibility of the filter cake strongly depends on the internal structure: a stabilized suspension leads to a dense packing even at low compressive loads, while the lose structure resulting from agglomerated particles can be easily compressed by rearrangement of the particles, so that the agglomerates can be considered as deformable particles. This effect is stronger for smaller particles. The compressibility is negligible at an ionic strength of 0.1 mol/L, where the particles are in very close contact and increases again for lower ionic strengths, where the nearest-neighbor particle distance reflects the balance between compression and DLVO interactions.

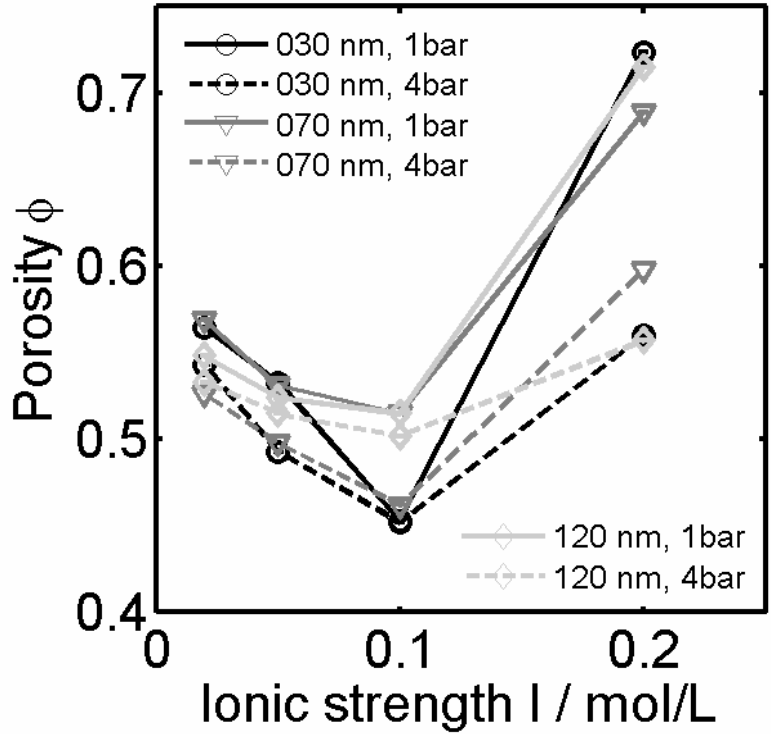

**Figure 6: Porosity of the filter cakes depending on the ionic strength of the suspension for different particle diameters and for different compressive loads.**

The porosity of the filter cakes also depends on the zeta potential of the particles (see figure 7). At low zeta potentials, the porosity is relatively high for all particle diameters and for all compressive loads because the particles are agglomerated, which is also observed in the experiments [61]. At a zeta potential of 40 mV, the Coulomb repulsion prevents agglomeration, causing a lower porosity. A further increase of the zeta potential increases the repulsion and thus the distance between the particles, in analogy to what happens when the ionic strength is decreased. Also the influence of the compression is analogous.

The error bars at a zeta potential of 0 mV indicate the 95% confidence intervals based on five simulations with different seeds for the random number generator, carried out for each particle diameter and each compressive load. These are the suspected worst cases, since the structures are more regular for stable suspensions.

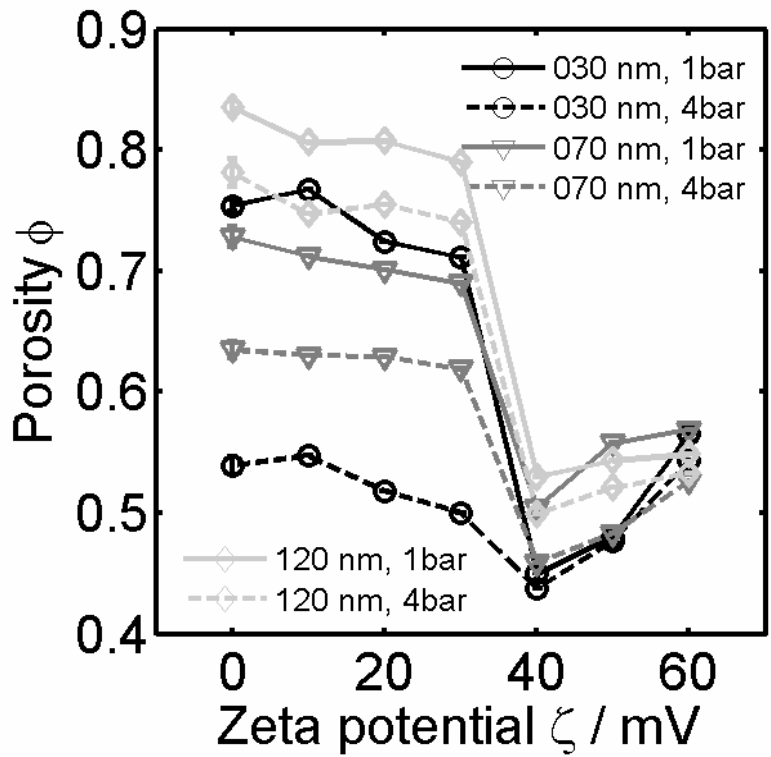

**Figure 7: Porosity of the filter cakes depending on the zeta potential of the particles for different particle diameters and for different compressive loads.**

### *6.3. Permeability*

Also the permeability of the filter cakes depends on the ionic strength of the suspension (see figure 8), which can be attributed to the changing pore size. The mean velocity in the pores decreases with decreasing pore diameter in the regime of laminar flow. The logarithmic plots of the permeability are similar in shape to the linear plot of the porosity. Upon an increase of the ionic strength, the permeability first decreases because the particles come closer due to the smaller Debye length. Above 0.1 mol/L, where the particles agglomerate, the permeability increases significantly. The permeability also decreases with increasing compression and with decreasing particle size for all ionic strengths. The pores between the particles are smaller for smaller particles if the porosity is identical.

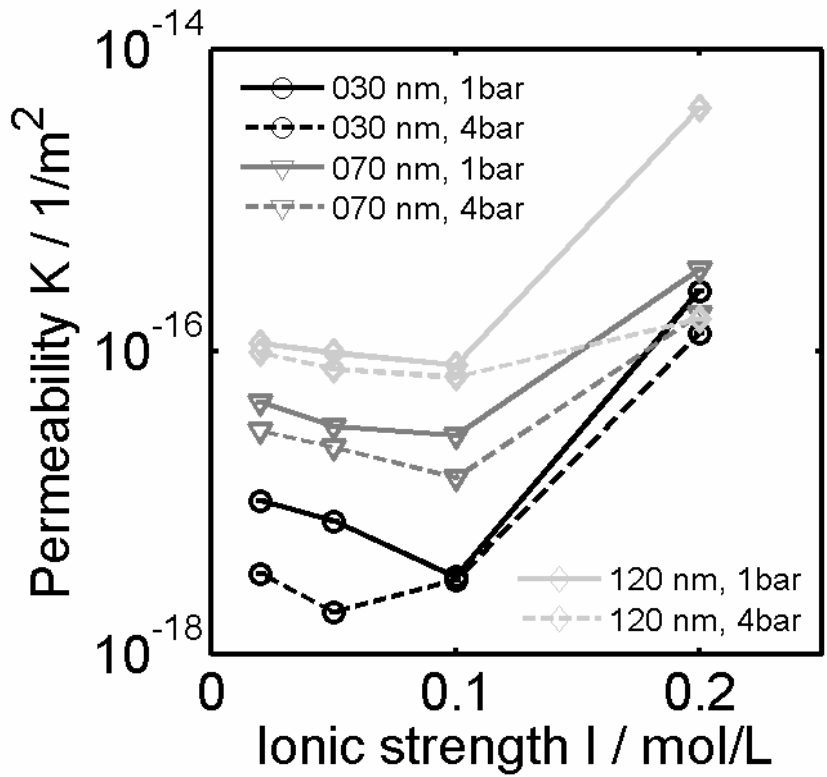

**Figure 8: Permeability of the filter cakes depending on the ionic strength of the suspension for different particle diameters and for different compressive loads.**

Figure 9 shows the permeability for a variation of the zeta potential. The permeability decreases with increasing compression and with decreasing particle size. Upon increasing the zeta potential up to 40 mV, the permeability decreases because of the decreasing porosity. Beyond 40 mV, the permeability increases again, especially for the smallest particles. This effect is also stronger for weaker compression.

Again, the error bars are shown for the suspected worst cases, which are the packed beds resulting from agglomerated particles with a zeta potential of 0 mV. The 95% confidence intervals are based on five simulations with different seeds for the random number generator, carried out for each particle diameter and each compressive load.

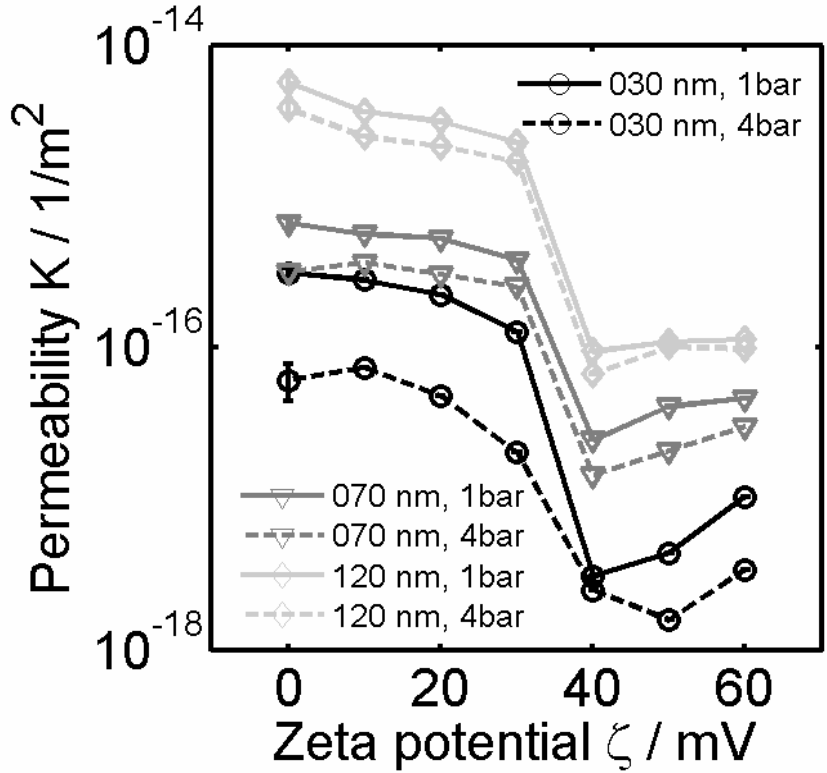

**Figure 9: Permeability of the filter cakes depending on the zeta potential of the particles for different particle diameters and for different compressive loads.**

The lattice Boltzmann (LB) simulations reveal that, for each particles size, the permeability of a packed bed is an exponential function of its porosity, which itself depends on the zeta potential, the ionic strength, and the compressive loads (see figure 10). A significant deviation from the exponential relation between the porosity and the permeability is found only for the 30 nm particles, where the simulated permeabilities are too high for porosities around 0.55. This indicates a stronger influence of the pore size heterogeneity on the permeability for smaller particles.

The simulated permeabilities can be approximated by an exponential function of the porosity

$$K = a \cdot e^{b \cdot \Phi}, \qquad (25)$$

with the coefficients $a$ and $b$ given in table 2 for the different particle sizes. The influence of the porosity, as given by the coefficient $b$, is similar for the different particle sizes, with values between 13.7 and 15.6. The parameter $a$ reflects the higher permeability of packed beds composed of larger particles.

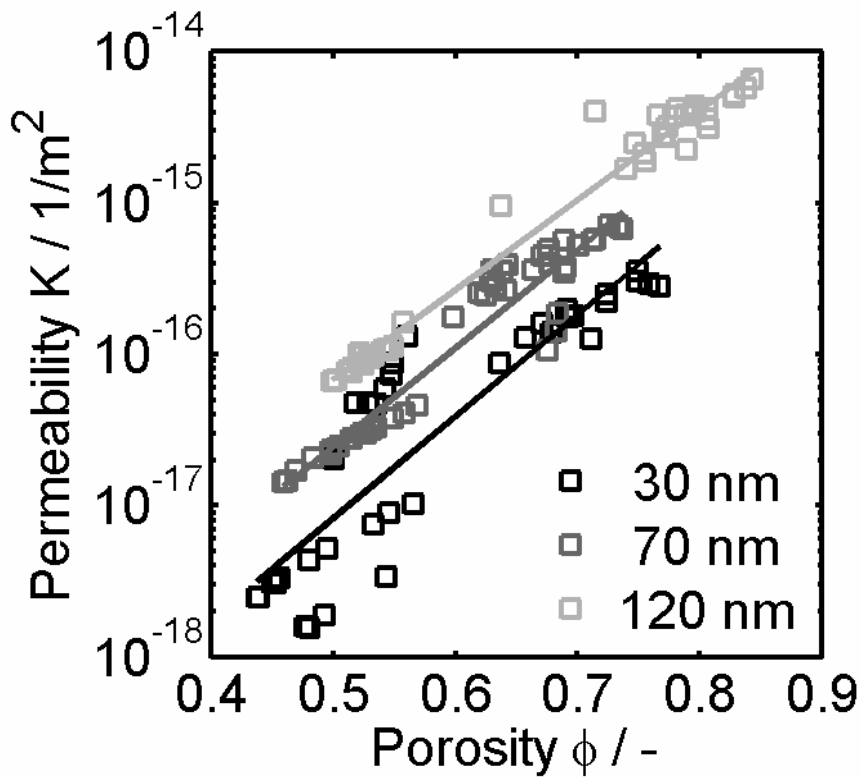

**Figure 10: Permeability depending on the porosity for different particle diameters as resulting from the simulations.**

Similar exponential relations between the porosity and the permeability are found in the experiments on the permeability of filter cakes consisting of colloidal boehmite particles (see figure 11). The influence of the porosity, which is contained in the parameter $b$, is again similar for the different particle sizes and the experiments are very close to the simulations. The deviation of the coefficient $a$ is caused by the polydispersity and the non-spherical form of the particles in the experiment. While for the monodisperse spheres in the simulation, the regimes of stable and agglomerates particles can be clearly separated, the irregular shape of the particles in the experiments smoothes the transition between the agglomerated and unagglomerated state.

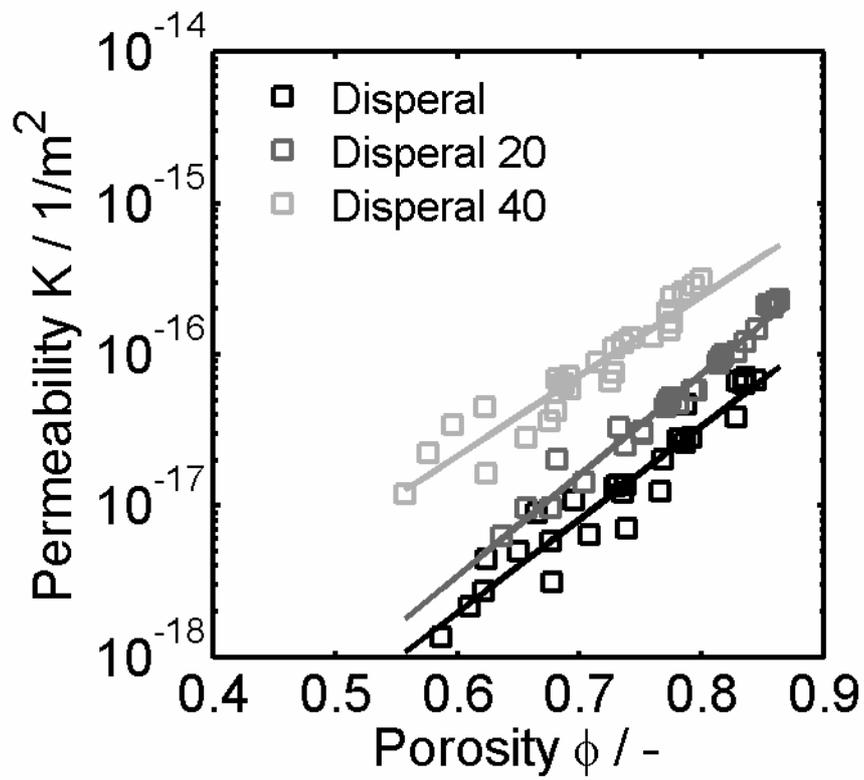

**Figure 11: Permeability depending on the porosity for different particle diameters as resulting from the experiments.**

**Table 2: Coefficients for the permeability as a function of the porosity**

| Particle diameter | Simulation | | Experiment | |
|---|---|---|---|---|
| | $a$ | $b$ | $a$ | $b$ |
| 30 nm | $3,3 \cdot 10^{-21}$ | 15,6 | $4.85 \cdot 10^{-22}$ | 13.9 |
| 70 nm | $2 \cdot 10^{-20}$ | 14,2 | $6 \cdot 10^{-22}$ | 14.7 |
| 120 nm | $7 \cdot 10^{-20}$ | 13.7 | $2.9 \cdot 10^{-20}$ | 11.3 |

# 7. Summary

This article presents a combination of molecular dynamics (MD) and stochastic rotation dynamics (SRD), which is used to simulate the agglomeration of colloidal particles as depending on the particles' zeta potential and the suspensions' ionic strength, as predicted by the DVLO theory. Analyzing the pair correlation function permits to determine the time for the agglomeration to reach equilibrium and to distinguish between primary and secondary agglomerates. The resulting suspensions are subsequently filtered by incrementally reducing the MD simulation space in order to form filter cakes. The porosities of these filter cakes depend on the compressive load and the agglomeration of the particles and agree qualitatively with experimental results for particles in the same size range. The permeabilities of the filter cakes, as determined by lattice Boltzmann (LB) simulations, reveal an exponential dependency on the porosity, which is corroborated by experimental investigations.

# 8. Acknowledgements

We thank the German Science Foundation (DFG) for funding within the priority program SPP 1164. The Jülich Supercomputing Center and the Scientific Supercomputing Center Karlsruhe are acknowledged for providing the required computing time.

# 9. Symbols (for the referee)

| | | | |
|---|---|---|---|
| $a$ | Coefficient | $k_B$ | Boltzmann constant |
| $A_C$ | Filter cake cross section area | $K$ | Permeability |
| $A_H$ | Hamaker constant | $K_{Hertz}$ | Hertz constant |
| $Ar$ | Archimedes number | $Kn$ | Knudsen number |
| $b$ | Coefficient | $L_z$ | Dimension of the simulation space in vertical direction |
| $Bo$ | Boltzmann number | | |
| $c_{Lub}$ | Lubrication constant | $L_{Cell}$ | Length of the SRD cells |
| $c_S$ | Speed of sound | $m$ | Mass |
| $d$ | Distance between the particle centers | $m_m$ | Mass of the particle $m$ |
| $d_{50,0}$ | Number-rated average particle diameter | $m_{SRD}$ | Mass of one SRD particle |
| | | $M$ | Molecular mass of the fluid |
| | | $Ma$ | Mach number |
| $d_P$ | Particle diameter | $N$ | Number of particles |
| $D$ | Diffusion constant | $N_k$ | Number of particles in SRD cell $k$ |
| $D_H$ | Hooke constant | $\langle N_k \rangle$ | Average number of SRD particles per cell |
| $e$ | Elementary charge | | |
| $\mathbf{e}_j$ | Lattice vectors | $N_A$ | Avogadro constant |
| $F$ | Faraday constant | $N_{MD}$ | Number of MD particles |
| $\mathbf{F}_m$ | Force on particle $m$ | $Pe$ | Péclet number |
| $F_{Lub}$ | Lubrication force | $R_C$ | Cake resistance |
| $f_j$ | Boltzmann distribution function for lattice vector $\mathbf{e}_j$ | $r_{CO,i}$ | Inner cut-off radius |
| | | $R_F$ | Flow resistance |
| $f_j^{eq}$ | Equilibrium distribution function for lattice vector $\mathbf{e}_j$ | $R_M$ | Membrane resistance |
| | | $Re$ | Reynolds number |
| $g$ | Gravitational constant | $Sc$ | Schmidt number |
| $G$ | Pair correlation function | $t$ | Time |
| $h_C$ | Thickness of the filter cake | $T$ | Temperature |
| $I$ | Ionic strength | $T_{SRD}$ | Temperature of the SRD fluid |
| | | $v_{LB}$ | Macroscopic velocity |
| $v_{Sed}$ | Stokes velocity | $\rho_{LB}$ | LB density |
| $V$ | Simulation volume | $\rho_S$ | Solid density |
| $\dot{V}_L$ | Volume flux | $\tau_{Diff}$ | Diffusion time |
| $w_j$ | Lattice weights | $\tau_L$ | Liquid relaxation time |
| $x_m$ | Position of particle $m$ | $\tau_{LB}$ | LB relaxation time |
| $x_{z,m}$ | Position of particle $m$ | $\tau_P$ | Particle relaxation time |
| $\dot{x}_m$ | Velocity of particle $m$ | $\tau_{Sed}$ | Sedimentation time |
| $\dot{\bar{x}}_k$ | Mean velocity of the particles in the cell $k$ | $\tau_{TF}$ | Thermal fluctuation time |
| | | $\Omega_{LB,j}$ | LB collision operator for lattice vector $e_j$ |
| $\ddot{x}_m$ | Acceleration of particle $m$ | | |
| $z$ | Ion valency | $\Omega_{SRD,k}$ | SRD rotation matrix for cell $k$ |
| Greek symbols | | | |
| $\Delta p$ | Pressure difference | $\Psi_{vdW}$ | Van-der-Waals potential |
| $\Delta t_{LB}$ | Time step | $\Psi_{Coul}$ | Coulomb potential |
| $\Delta t_{SRD}$ | Time step | $\Psi_{Hertz}$ | Hertz potential |
| $\varepsilon_0$ | Vacuum permittivity | $\zeta$ | Zeta potential |
| $\varepsilon_{rel}$ | Relative permittivity | Indices | |
| $\Phi$ | Porosity | i | Index |
| $\varphi_j$ | Additional forces term in LB | j | Index for lattice vectors |
| $\eta$ | Dynamic viscosity of the permeate | k | Index for cells |
| $\eta_{SRD}$ | Dynamic viscosity of the SRD fluid | m | Index for particles |
| $\kappa$ | Reciprocal Debye length | n | Index for particles |
| $\lambda$ | Mean free path of the SRD particles | C | Cake |
| $\rho°$ | Reference density | L | liquid |
| $\rho_L$ | Liquid density | S | solid |